\documentclass[hyper]{JHEP} % 10pt is ignored!

\usepackage{epsfig}
\usepackage{graphicx}% Include figure files

\def\be{\begin{equation}}
\def\ee{\end{equation}}
\def\ba{\begin{array}}
\def\ea{\end{array}}
\def\bea{\begin{eqnarray}}
\def\eea{\end{eqnarray}}
\def\nn{\nonumber\\}

\def\ct{\cite}
\def\la{\label}
\def\eq#1{Eq. (\ref{#1})}

\def\G{\Gamma}

\def\D{\Delta}

\def\ph{\phi}

\def\ps{\psi}

\def\k{\kappa}

\def\n{\nu}
\def\th{\theta}

\def\r{\rho}
\def\s{\sigma}

\def\ta{\tau}
\def\o{\omega}

\def\pa{\partial}

\def\lb{\left[}

\def\ls{\left(}

\def\rb{\right]}

\def\rs{\right)}
\def\td{\tilde}

\def\det{{\rm det}}

\newcommand\fverb{\setbox\pippobox=\hbox\bgroup\verb}

\newcommand\fverbdo{\egroup\medskip\noindent%

            \fbox{\unhbox\pippobox}\ }

\newcommand\fverbit{\egroup\item[\fbox{\unhbox\pippobox}]}

\newbox\pippobox

\title{Semiclassical strings in $AdS_3 \times S^2$}

\author{Bogeun Gwak$^a$, Bum-Hoon Lee$^{ab}$, Kamal L. Panigrahi$^c$ and Chanyong Park$^d$\\

\vspace{1cm}

$^a$Department of Physics, Sogang University, Seoul 121-742, Korea \\

$^b$ Center for Quantum Spacetime (CQUeST), Sogang University, Seoul 121-742, Korea \\

$^c$ Department of Physics, Indian Institute of Technology
Guwahati, Guwahati-781 039, India \\

$^d$ National Institute for Mathematical Sciences, 385-16 Doryong-dong, Yuseong-gu, Daejeon 305-340, Korea  \\

\vspace{1cm}

E-mail: \email{rasenis@sogang.ac.kr, bhl@sogang.ac.kr,
panigrahi@iitg.ernet.in, cyong21@nims.re.kr}}

\preprint{}

\abstract{ In this paper, we investigate the semiclassical strings
in $AdS_3 \times S^2$, in which the string configuration of
$AdS_3$ is classified to three cases depending on the parameters.
Each of these has a different anomalous dimension proportional to
$\log S$, $S^{1/3}$ and $S$, where $S$ is a angular momentum on
$AdS_3$. Further we generalize the dispersion relations for
various string configuration on $AdS_3 \times S^2$. }


\begin{document}
%%%%%%%%%%%%%%%%%%%%%%%%%%%%%%%%%%%%%%%%%%
%%%%redukovana verze clanku            %%%%
%%%%o Ispike_a3.tex%%%%%%%%%%%%%%%%%%%%%%%
%%%%%%%%%%%%%%%%%%%%%
%%%%Introduction %%%%%%%%%
%%%%%%%%%%%%%%%%%%%%

\section{Introduction}\label{first}

A remarkable development in the study of string theory of last
decade is the celebrated string theory-gauge theory duality
\cite{Maldacena:1997re}, which relates the spectrum of
semiclassical string states on $AdS_5 \times S^5$ to the operator
dimensions of ${\cal N}=4$ supersymmetric Yang-Mills (SYM) theory
in four dimensions\cite{Gubser:1998bc,Witten:1998qj}. This
state-operator matching has been achieved in the planar limit,
which simplifies considerably both sides of the duality. Recently
another example of gauge theory-string theory duality has been
proposed based on the worldvolume dynamics of multiple M2-branes.
This relates type IIA string theory in $AdS_4 \times CP^3$ with
${\cal N}=6$ Chern-Simons matter theory in three dimensions
\cite{Aharony:2008ug}\footnote{for related interesting work see
\cite{Nishioka:2008gz,Minahan:2008hf,Gaiotto:2008cg,
Arutyunov:2008if,Gomis:2008jt,Stefanski:2008ik,Bak:2008cp}.}.

In the study of the gauge/gravity duality, an interesting
observation is that the ${\cal N}=4$ SYM theory can be described
by the integrable spin chain model
\ct{Minahan:2002ve,Beisert:2003yb,Beisert:2004ry}. It was further
noticed that the string theory also has as integrable structure in
the semiclassical limit. This integrability nature of both the
gauge theory and string theory side has been used extensively to
understand the duality better. In this connection, Hofman and
Maldacena (HM) \ct{hofman} considered a special limit where the
problem of determining the spectrum of both sides becomes rather
simple. The spectrum consists of an elementary excitation known as
magnon which propagate with a conserved momentum $p$ along the
long spin chain. In the dual formulation, the most important
ingredient is semiclassical string solutions, which can be mapped
to long trace operator with large energy and large angular
momenta. For some related work see for example
\cite{Dorey:2006dq}-\cite{Kluson:2008gf}.

Not so long back the giant magnon and spike solutions for the
string on $S^2$ or $S^3$ have been studied
\ct{Ishizeki:2007we,Chen:2006gea,Lee:2008yq}. In addition, the solitonic
string configuration in the $AdS$ space, whose anomalous dimension
corresponds to that of the twist two operator of SYM theory or the
cusp anomaly, was considered
\ct{Gubser:2002tv,Alday:2007he,Alday:2008cg,Kruczenski:2007cy}.
More recently the semiclassical string configuration on $AdS_3
\times S^3$ has also been investigated in a special parameter
region \ct{Lee:2008sk,Ryang:2006yq}. However a general class of
solutions for string moving in $AdS_3 \times S^3$ background has
been lacking. The aim of this paper is to classify such solutions
with the hope that a better understanding of these will enable us
to investigate more complicated solutions of the dual theory on
the boundary.

In this paper, we consider the semiclassical strings moving in
$AdS_3 \times S^2$ background and investigate the relation among
various conserved charges of the string configuration at various
parameter regions. Depending on the parameter region, the
macroscopic strings on $AdS_3$ can be classified to three cases,
whose anomalous dimension is proportional to $\log S$, $S^{1/3}$
or $S$, where $S$ is a conserved angular momentum on $AdS_3$.
After that, we generalize rotating string on $AdS_3$ to the one
rotating on $AdS_3 \times S^2$ and study the corresponding
dispersion relation or the anomalous dimension.

The rest of the paper is organized as follows. In the section 2,
we write down the most general equations describing the rotating
string in $AdS_3 \times S^2$ background. In the section 3, first
we classify various possible string configuration in $AdS_3$.
Depending on the region of parameter there are three distinct
solutions with different anomalous dimensions. We further
generalize them further by looking at the various shapes of the
string. We write down the relevant dispersion relations in all the
cases. Section 4 is devoted to the study of more general string
configurations in $AdS_3 \times S^2$ and we find out the
dispersion relations among various conserved charges. In section
5, we present our conclusions.

\section{Semiclassical String configurations in $AdS_3 \times S^2$}
In this section we will study the general rotating string solution
in $AdS_3 \times S^2$. We start by writing down the relevant
metric for $AdS_3 \times S^2$ in a particular coordinate system,
\begin{equation}        \la{action}
ds^2=\frac{1}{4}R^2[-\cosh^2\rho dt^2+d\rho^2+\sinh^2\rho
d\phi^2+d\theta^2+\sin^2\theta d\psi^2] .
\end{equation}
The Polyakov action for the string moving in this background can be written as \\
\begin{equation}
S=\frac{1}{4\pi}\int d^2\sigma\sqrt{-\det\,h}h^{\alpha\beta}\partial_\alpha
x^\mu\partial_\beta x^\nu G_{\mu\nu}\,.
\end{equation}
We choose the following parametrization for the semiclassical
rotating string in this background
\begin{eqnarray}        \la{parameter}
t &=& k\tau+h_1(y) , ~~~\rho = \rho(y) , ~~~\phi =
\omega\tau+h_2(y) , \nonumber \\ \theta &=& \theta(y) , ~~~\psi =
\nu\tau+g(y) ,
\end{eqnarray}
where $y = a \ta + b \s$. The relevant equations of motion reduce
to
%Due to the rotational and translational symmetries in $\ph$,
%$\ps$ and $t$, the equations of motion for them are reduced to the
%first order differential equations
\begin{eqnarray}
h_1'&=&\frac{1}{b^2-a^2}(ak - \frac{\pi_0}{\cosh^2\rho})  ,\nonumber\\
h_2'&=&\frac{1}{b^2-a^2}(a\omega - \frac{\pi_2}{\sinh^2\rho})  ,\nonumber\\
g'&=& \frac{1}{b^2-a^2}(a\nu - \frac{\pi_4}{\sin^2\th}) ,
\end{eqnarray}
where $\pi_0$, $\pi_2$ and $\pi_4$ are integration constants. The
above equations of motion have to be supplemented by the Virasoro
constraints, which are simply given by, in a particular form \bea
0 &=&
T_{\tau\tau}+T_{\sigma\sigma}-\frac{a^2+b^2}{ab}T_{\tau\sigma} ,
\nn 0 &=& T_{\tau\tau}+T_{\sigma\sigma}-2T_{\tau\sigma} . \eea The
first Virasoro constraint gives rise to a relation among various
parameters \be \la{pi4con} \n \pi_4  = k\pi_0 - \omega\pi_2 . \ee
The second one becomes
\begin{eqnarray}
0 &=& - \frac{1}{(b^2-a^2)^2}(b^2k^2+2bk\pi_0)\nonumber\\
&& + \rho'^2 +\frac{1}{(b^2-a^2)^2} \ls 2b\omega\pi_2+b^2(\omega^2-k^2)\sinh^2\rho
+\frac{(\pi_2^2-\pi_0^2)\sinh^2\rho+\pi_2^2}{\cosh^2\rho \sinh^2\rho} \rs \nonumber\\
&& +\theta'^2 +\frac{1}{(b^2-a^2)^2}\ls
\nu^2b^2\sin^2\theta+2b\nu\pi_4+\frac{\pi_4^2}{\sin^2\theta} \rs .
\nonumber \\ &=& - k_0^2 + k_1^2+k_s^2,
\end{eqnarray} where in the last equality we have defined
\bea
k_0^2 &=&  \frac{1}{(b^2-a^2)^2}(b^2k^2+2bk\pi_0) , \nonumber\\
k_1^2 &=&  \rho'^2 +\frac{1}{(b^2-a^2)^2} \ls 2b\omega\pi_2+b^2(\omega^2-k^2)\sinh^2\rho
+\frac{(\pi_2^2-\pi_0^2)\sinh^2\rho+\pi_2^2}{\cosh^2\rho \sinh^2\rho} \rs ,\nonumber\\
k_s^2&=& \theta'^2 +\frac{1}{(b^2-a^2)^2}\ls
\nu^2b^2\sin^2\theta+2b\nu\pi_4+\frac{\pi_4^2}{\sin^2\theta} \rs .
\eea Then, the differential equations for $\r$ and $\th$ can be
rewritten as
\begin{eqnarray}
\rho'^2 &=& \frac{1}{(b^2-a^2) \cosh\rho \sinh\rho} \Big{[} b^2(\omega^2-k^2) \sinh^6\rho
+\Big{(} 2b\omega\pi_2+b^2(\omega^2-k^2)  \nonumber \\
&& + (b^2-a^2)k_1^2 \Big{)} \sinh^4\rho +\Big{(}
2b\omega\pi_2+(b^2-a^2)^2k_1^2+(\pi_2^2-\pi_0^2) \Big{)}
\sinh^2\rho+\pi_2^2 \Big{]} ,\nn \theta '^2
&=&\frac{1}{(b^2-a^2)^2} \frac{\nu^2 b^2}{\sin^2\theta} \lb
-\sin^4\theta + \ls \frac{(b^2-a^2)^2
k_s^2}{b^2\nu^2}-\frac{2\pi_4}{b\nu} \rs \sin^2\theta
-\frac{\pi_4^2}{b^2 \nu^2} \rb. \eea The equation for $\th$ can be
rewritten as \be \th' = \frac{b\n}{(b^2-a^2) \sin \th}
\sqrt{(\sin^2 \th_{max} - \sin^2 \th) ( \sin^2 \th - \sin^2
\th_{min})} , \ee for $k_s^2 (b^2-a^2)^2 > 2 b\n \pi_4$ only. The
condition $\sin \th_{max} = 1$ for the infinite size magnon or
spike gives a relation \be k_s = \frac{b \n + \pi_4}{b^2-a^2} .
\ee Using this, $\sin \th_{min}$ becomes \be \sin \th_{min} =
\frac{\pi_4}{b \n} , \ee where $b\n > \pi_4$.

\section{Classification of the string configurations on $AdS_3 \times S^2$}

From now on, we will consider the special case, $k = \o >  0$
only. An example for $k \ne \o$ was investigated in Ref.
\ct{Lee:2008sk}. For $\k = \o$, the equations of the $AdS$ part
are reduced to
\begin{eqnarray}        \la{class}
\rho' &=& \frac{1}{(b^2 - a^2) \cosh \rho \sinh \rho}
\sqrt{ \Delta \sinh^4\rho+\G \sinh^2\rho-\pi_2^2  }, \nn
\ph' &=& \frac{1}{(b^2 - a^2) }  \ls  a \o  - \frac{\pi_2}{\sinh^2 \r}  \rs ,
\eea
with
\bea        \la{recl}
&& \G = \Delta + \pi_0^2 -\pi_2^2 ,  \nn
&&\Delta = (a^2-b^2)^2 k_1^2 -2b\omega\pi_2 .
\end{eqnarray}
Since there exists a solution only when the inside of the square
root in the first equation of \eq{class} is positive, we will
investigate the range of $\r$ accordingly. Depending on the
parameters, the range of $\r$ can be classified as the follows: \\

\noindent  I. $\D > 0$

In this case, there exists one boundary value $\r_{min}$ so that
the range of $\r$ is given by $\r_{min} \le \r < \r_{max} =
\infty$ and $\r'$ is rewritten as \be \r' = \frac{\sqrt{\D}}{b^2-
a^2} \frac{\sqrt{\sinh^2 \r + \sinh^2 \r_0}}{\sinh \r \ \cosh \r}
\sqrt{\sinh^2 \r - \sinh^2 \r_{min} } , \ee where depending on the
sign of $\G$, on has the following choices
%$\sinh^2 \r_0$ and
%$\sinh^2 \r_{min}$ are given by
\bea && {\rm i)} \ \G > 0  \nn && \qquad \qquad \sinh^2 \r_0 =
\frac{ \sqrt{\G^2 + 4 \pi_2^2 \D }+\G}{2 \D} \nn
&& \qquad \qquad  \sinh^2 \r_{min} = \frac{\sqrt{\G^2 + 4 \pi_2^2 \D } - \G }{2 \D} \ , \\
&& {\rm ii)} \ \G = 0  \nn
&& \qquad \qquad  \sinh^2 \r_0 = \sinh^2 \r_{min} = \frac{\sqrt{\D+\pi_0^2}}{ \sqrt{\D}}  , \\
&& {\rm iii)} \ \G <  0  \nn && \qquad \qquad  \sinh^2 \r_0 =
\frac{\sqrt{|\G|^2 + 4 \pi_2^2 \D } - |\G|}{2 \D}   \nn && \qquad
\qquad   \sinh^2 \r_{min} = \frac{\sqrt{|\G|^2 + 4 \pi_2^2 \D } +
|\G| }{2 \D} . \eea

In addition, the string configuration can also be classified into
various cases by looking at the shape of the string. For this, we
define the slope of the string at a fixed $\ta$ as \be {\cal R}
\equiv \left| \frac{\pa \r}{\pa \ph} \right| = \left|
\frac{\r'}{\ph'} \right| \  . \ee For example, the string slope is
infinity, and the string configuration has a cusp at $\r_{min}$.
For more details, we introduce $\sinh^2 \r_c = \frac{\pi_2}{a \o}$
satisfying $\ph' = 0$ in which the slope ${\cal R}$ diverges.

{\bf case (i)} If $\r_c < \r_{min}$, the string slope becomes $0$
at $\r_{min}$ and a constant at $\r_{max} = \infty$.

{\bf case (ii)} In the case of $\r_c = \r_{min}$, as previously
mentioned, the slope at $\r_{min}$ becomes infinity so that this
string configuration has a cusp at $\r_{min}$ and a constant slope
at $\r_{max}=\infty$.

{\bf case (iii)} If $\r_c > \r_{min}$ but finite, the string
configuration is same as case (i) except that it includes a point
$\r_c$ where the sign of the slope is opposite.

{\bf case (iv)} If $\r_c = \r_{max} = \infty$, the slope becomes
zero at $\r_{min}$ and infinity at $\r_{max}$. This case can be obtained
when we consider $a=0$. \\

\noindent  II. $\D = 0$

In this case, the range of $\r$ is given by $\r_{min} \le \r <
\infty$. For $\G > 0$ $\r'$ becomes \be \r' =
\frac{\sqrt{\G}}{b^2- a^2} \frac{\sqrt{\sinh^2 \r - \sinh^2
\r_{min} }}{\sinh \r \ \cosh \r}  , \ee with \be \sinh^2 \r_{min}
= \frac{\pi_0^2}{\pi_0^2 - \pi_2^2} . \ee  The classification of
the string configuration is similar to the previous case. Note
that since $\r'$ becomes zero at $\r=\infty$, the string
configuration is slightly different from the previous cases. For
$\r_c < \r_{min}$, the slope ${\cal R}$ vanishes at $\r_{min}$ and
$\r_{max}=\infty$. For $\r_c = \r_{min}$, the string slope becomes
infinity at  $\r_{min}$ and zero at $\r_{max}=\infty$. For $\r_c
> \r_{min}$ but finite, ${\cal R}$ becomes zero at both
$\r_{min}$ and $\r_{max}=\infty$. For $\r_c  = \r_{miax}$, with
$a=0$, ${\cal R}$ is zero at $\r_{min}$ and infinity at $\r_{max}=\infty$. \\

In the case of $\G \le 0$, there is no string configuration since
the inside of the square root in the first equation of \eq{class}
becomes negative and hence $\r'$ becomes imaginary.

\noindent  III. $\D < 0$

In this case, there exist two boundary values for $\G > 0$ so that
the range of $\r$ is given by $\r_{min} \le \r \le \r_{max}$.
Then, $\r'$ is \be \r' = \frac{\sqrt{|\D|}}{b^2- a^2} \frac{\sqrt{
\ls \sinh^2 \r_{max} - \sinh^2 \r \rs \ls \sinh^2 \r - \sinh^2
\r_{min} \rs }}{\sinh \r \ \cosh \r} , \ee where $\sinh^2
\r_{max}$ and $\sinh^2 \r_{min}$ are \bea \sinh^2 \r_{max} &=&
\frac{\G + \sqrt{\G^2 - 4 \pi_2^2 |\D|} }{2 |\D|}  \ ,\nn \sinh^2
\r_{min} &=&  \frac{\G - \sqrt{\G^2 - 4 \pi_2^2 |\D|} }{2 |\D|} .
\eea For $\r_c < \r_{min}$ or $\r_c > \r_{max}$, the slope
vanishes at $\r_{min}$ and $\r_{max}$. Unlike the previous cases,
the case of $a=0$ corresponds to the case of $\r_c > \r_{max}$.
For $\r_c = \r_{min}$, the string configuration has a cusp at
$\r_{min}$ where ${\cal R}$ becomes infinite and a zero slope at
$\r_{max}$. For $\r_{min} < \r_c < \r_{max}$, the slopes at
$\r_{min}$ and $\r_{max}$ vanish and there exist a point where the
sign of slope is changed. For $\r_c = \r_{max}$, the slope
vanishes  at $\r_{min}$ and becomes infinite at $\r_{max}$. Once
again for $\G \le 0$, there is no string configuration because
$\r'$ becomes imaginary.

In the infinite size limit for the giant magnon or spike solutions
for the string on $S^2$ with $\th_{max} = \pi/2$, the conserved
charges are given by:
\begin{eqnarray}    \la{conserved ch}
E &=& \left| \frac{2T}{(b^2 - a^2)b} \int_{\r_{min}}^{\r_{max}} d \r
\frac{(b^2 \o \cosh^2 \rho - a \pi_0)}{\r'} \right| , \nn
S &=& \left| \frac{2T}{(b^2 - a^2)b} \int_{\r_{min}}^{\r_{max}} d \r
\frac{(b^2 \omega \sinh^2\rho - a \pi_2)}{\r'}  \right| , \nn
J &=& \left| \frac{2T}{(b^2 - a^2)b} \int_{\th_{min}}^{\pi/2}
d \theta \frac{(b^2 \nu \sin^2\theta - a \pi_4)}{\theta'} \right|  , \nn
\Delta\phi &=& \left| \frac{2}{(b^2 - a^2)}\int_{\r_{min}}^{\r_{max}}
d\r \frac{1}{\r'}(a \omega - \frac{\pi_2}{\sinh^2\rho})  \right| , \nn
\Delta\psi &=& \left| \frac{2}{(b^2 - a^2)}\int_{\th_{min}}^{\pi/2}
d\theta \frac{1}{\theta'}(a \nu - \frac{\pi_4} {\sin^2\theta})  \right| ,
\end{eqnarray}
where we consider only the positive quantities. For simplicity, we
assume that the ends of the open string are located at $\r_{max}$
in the AdS space and at the same time, $\th_{max}$ on $S^2$.

\subsection{Spike on  $AdS_3$ and a point-like string on $S^2$}

At first, we consider the string configuration located at a point
on $S^2$ ($\th'=0$). For this, we choose $\n=\pi_4=0$, which makes
the angular momentum and the angle difference in the $\ps$
directions to vanish. So, the string solution for $\n=\pi_4=0$
corresponds to the string extend in $AdS$ but a static point
particle on $S^2$. Once again we consider the situation case by
case like the previous section for different parameter regions. \\

\noindent I. $\D >0$

\noindent 1) We first consider the case with $a \ne 0$ and $b^2 \o
\ne a (\pi_0-\pi_2)$. The interesting physical quantities of this
string configuration are \bea E - S &=& \frac{2 \ls b^2 \o - a
(\pi_0 - \pi_2) \rs T}{b \sqrt{\D}} \int_{\r_{min}}^{\r_{max}} d
\r \frac{\sinh \r \ \cosh \r}{\sqrt{\ls \sinh^2 \r + \sinh^2 \r_0
\rs  \ls \sinh^2 \r - \sinh^2 \r_{min} \rs}}  , \nn S &=&
\frac{2T}{b \sqrt{\D}} \int_{\r_{min}}^{\r_{max}} d \r \frac{\sinh
\r \ \cosh \r \ls b^2 \o \sinh^2 \r -  a \pi_2 \rs}{\sqrt{\ls
\sinh^2 \r + \sinh^2 \r_0 \rs  \ls \sinh^2 \r - \sinh^2 \r_{min}
\rs}}  , \nn \Delta\phi &=&\frac{2}{\sqrt{\D}}
\int_{\r_{min}}^{\r_{max}} d\r \frac{\cosh \r \ls a \o \sinh^2 \r
-  \pi_2 \rs }{\sinh \r \sqrt{\ls \sinh^2 \r + \sinh^2 \r_0 \rs
\ls \sinh^2 \r - \sinh^2 \r_{min} \rs}} , \eea where $\r_{max} =
\infty$. In the case of $\sinh \r_{min} >> 1$, the dispersion
relation can be approximately written as \be E - S \approx \frac{
b^2 \o - a (\pi_0 - \pi_2) }{a b \o } T  \D \ph . \ee Note that
the dominant contribution of the above integral comes from the
large $\r$ region. So the calculation of the integral gives rise
to \bea E-S &\approx& \frac{2 \ls b^2 \o - a (\pi_0 - \pi_2) \rs
T}{b \sqrt{\D}}   \r_{max} , \nn S &\approx& \frac{b \o T}{4
\sqrt{\D}} e^{2 \r_{max}} . \eea As a result, the dispersion
relation for $\r_{max} = \infty$ becomes \be E - S \approx
\frac{\ls b^2 \o - a (\pi_0 - \pi_2) \rs T}{b \sqrt{\D}} \log \ls
\frac{4 \sqrt{\D}}{b \o T} S \rs  , \ee
which is the generalization of the GKP string configuration \ct{Gubser:2002tv}. \\

\noindent 2) Let us consider the case $b^2 \o = a (\pi_0-\pi_2)$,
the dispersion relation is \be E - S = 0 , \ee which looks like
that of the BPS vacuum state. In this case, $S$ and $\D \ph$ are
given by \bea S &\approx& \frac{b \o T}{4 \sqrt{\D}} \lb e^{2
\r_{max}} + 4 \ls \sinh^2 \r_{min} - \sinh^2 \r_0 -  \frac{2 a
\pi_2}{b^2 \o } \rs \r_{max} \rb , \nn \D \ph &\approx& \frac{2 a
\o}{\sqrt{\D}} \r_{max} . \eea So $S$ and $E$ can be rewritten in
terms of $\D \ph$ as \be E = S \approx \frac{b \o T}{4 \sqrt{\D}}
\lb e^{\D \ph \sqrt{\D}  / a \o} - \ls e^{2\r_0} - e^{2\r_{min}} +
\frac{2 a \pi_2}{b^2 \o } \rs \frac{\D \ph  \sqrt{\D} }{ 2 a \o
}\rb .
\ee \\

\noindent 3) Now, consider the case $a=0$. In this case, $\D \ph$
has a finite value because the integral is regular even at
$\r_{min}$ and $\r_{max}=\infty$. On the contrary, $E-S$ and $S$
diverge at $\r_{max}=\infty$ so that we can not represent $E-S$
and $S$ in terms of $\D \ph$. However, we can compute the
dispersion relation for this string configuration, which is \be E
- S \approx \frac{ b \o T}{\sqrt{\D}}  \log \ls \frac{4
\sqrt{\D}}{b \o T} S \rs .
%+ \ls  e^{2\r_0} - e^{2\r_{min}}  \rs \frac{b \o T}{4 \sqrt{\D}}\rb .
\ee
Notice the logarithmic behavior of the anomalous dimension. \\
\noindent II. $\D=0$

For this case, $E-S$ and $S$ are given by
\bea
E - S &\approx& \frac{\ls b^2 \o - a (\pi_0 - \pi_2) \rs T}{b \sqrt{\G}} e^{ \r_{max}} , \nn
S &\approx& \frac{b \o T}{12 \sqrt{\G}} e^{3 \r_{max}} ,
\eea
so that the dispersion relation becomes
\be
E - S \approx \frac{\ls b^2 \o - a (\pi_0 - \pi_2) \rs T}{b \sqrt{\G}}
\ls \frac{12 \sqrt{\G}}{b \o T}  \rs^{1/3} S^{1/3} .
%- \half e^{2 \r_{min}} \ls \frac{b \o T}{12 \sqrt{\G} S} \rs^{1/3} \rb.
\ee Like the previous case, if we put $ b^2 \o - a (\pi_0 - \pi_2)
=0$, the string configuration becomes BPS-like configuration.
%In
%this case, the relation between $E$, $S$ and $\D \ph$ is given by
%\be E = S \approx \frac{b \G T}{12 a^3 \o^2} \D \ph^3 . \ee
In this case, the dispersion relation  becomes \be E - S \approx
12^{1/3} \ls \frac{ b \o  T}{ \sqrt{\G}} \rs^{2/3}  S^{1/3} .
\ee \\

\noindent III. $\D < 0$

For this case, the conserved charges are given by \bea E - S &=&
\frac{\ls b^2 \o - a (\pi_0 - \pi_2) \rs T}{b \sqrt{\left| \D
\right|}}  \pi , \nn S &=& \frac{T}{ b \sqrt{\left| \D \right|}}
\ls \frac{b^2 \o}{2} (\sinh^2 \r_{max} + \sinh^2 \r_{min}) - a
\pi_2 \rs  \pi . \eea From these, the dispersion relation  can be
rewritten as \be E - S =\frac{2 \ls b^2 \o - a (\pi_0 - \pi_2) \rs
\left| \D \right|}{b^2 \o \G -  2 a \pi_2 \left| \D \right|} S .
\ee For $a=0$,  the dispersion relation is reduced to a simple
form \be E - S =\frac{2 \left| \D \right|}{ \G } S . \ee

\subsection{Spike on $AdS_3$ and circular string on $S^2$}

Here, we consider the following parameter region: $\n=0$ and
$\pi_4 \ne 0$. As previously mentioned, $\n=0$ implies $\th'=0$.
So the string configuration in this parameter region describes a
circular string on $S^2$, which is extended in the $\ps$-direction
with an angular momentum $J$ at the fixed $\th=\th_c$. To describe
the angular momentum $J$ and the angle difference $\D \ps$, $\th$
is not a good variable so that using \be     \la{chrel} \frac{d
\th}{d \r} = \frac{\th'}{\r'} , \ee we can change the integral
with respect to $\th$ in \eq{conserved ch} to \bea J &=& \left| -
\frac{2T}{(b^2 - a^2)b} \int_{\r_{min}}^{\r_{max}} d \r \frac{ a
\pi_4 }{\r'} \right|  , \nn \Delta\psi &=& \left| - \frac{2}{(b^2
- a^2)}\int_{\r_{min}}^{\r_{max}} d\r \frac{\pi_4 }{\r'
\sin^2\th_c} \right| . \eea From these, we find \be J = \frac{a \
\sin^2 \th_c}{b} T \D \ps . \ee Especially, since the angular
momentum becomes zero for $a=0$, the string solution is reduced to
a static circular string on $S^2$ .

For $\D >0$, the dispersion relation is slightly modified to
\be
E - S - J \approx \frac{\ls b^2 \o - a (\pi_0 - \pi_2 +\pi_4) \rs T}{b \sqrt{\D}}
\log \ls \frac{4 \sqrt{\D}}{b \o T} S \rs .
\ee
So the BPS-like configuration appears at $ b^2 \o = a (\pi_0 - \pi_2 +\pi_4) $.
The dispersion relation for $a=0$ can be represented in terms of
the angle difference $\D \ps$ on $S^2$ instead of $\D \ph$ on $AdS$
\bea
E - S &=& \frac{b \o \sin^2 \th_c}{\pi_4} T \D \ps  \nn
         &\approx&  \frac{ b \o T}{\sqrt{\D}}  \log \ls \frac{4 \sqrt{\D}}{b \o T} S \rs .
\eea

For $\D = 0$, when $a \ne 0$ the modified dispersion relation is
\be
E - S - J \approx \frac{\ls b^2 \o - a (\pi_0 - \pi_2 +\pi_4) \rs T}{b \sqrt{\G}}
\ls \frac{12 \sqrt{\G}}{b \o T} \rs^{1/3}   S^{1/3} ,
\ee
and at $a=0$  it becomes
\be
E - S \approx 12^{1/3} \ls \frac{b \o  T}{ \sqrt{\G}} \rs^{2/3}    S^{1/3}
\ee

In the case of $\D < 0$,  the dispersion relation for $a \ne 0$ is
modified to \be E - S -J =\frac{2 \ls b^2 \o - a (\pi_0 - \pi_2
+\pi_4) \rs \left| \D \right|}{b^2 \o \G -  2 a \pi_2  \left| \D
\right|} S , \ee and one for $a=0$ becomes \be E - S  =\frac{2
\left| \D \right|}{\G} S . \ee As shown in the above results, at
$a=0$ the dispersion relations for a circular string on $S^2$ are
same as ones for a point-like string on $S^2$.

\subsection{Circular string on $AdS_3$ and magnon on $S^2$}

In this section, we consider the string configuration located at
$\r = \r_c$. Especially, for $\r_c = \r_{min} = 0$ and $\pi_2 = 0$
the string configuration on $AdS_3 \times S^2$ reduces to the
magnon or spike on $R \times S^2$ located at the center of
$AdS_3$.

For $\r_c \ne 0$, the string configuration on the $AdS_3$ part
describes a circular string wounded in the $\ph$-direction. The
dispersion relation for this configuration is given by \be
\la{discirs2} E - S - J = \frac{2T}{(b^2 - a^2)b}
\int_{\th_{min}}^{\pi/2} d \theta \frac{ b^2 \o -
a(\pi_0-\pi_2-\pi_4) - b^2 \nu \sin^2\theta }{\theta'} , \ee where
\eq{chrel} is used. To obtain magnon like solution, we impose \bea
b^2 \n &=& b^2 \o - a(\pi_0-\pi_2-\pi_4) , \nn a \n &=& \pi_4 ,
\eea such that the values of $E-S-J$ and $\D \ps$ are finite.
Using \eq{pi4con}, one gets \be \ls b^2 - \frac{a^2 \n^2}{\o^2}
\rs \o = (b^2 -a^2) \n . \ee In this case, the dispersion relation
is given by a similar form of a magnon on $R \times S^2$ \be E -S
-J = 2 T \sin \frac{p}{2} , \ee where $p = \D \ps$. In more
general cases, the conserved charges are represented in terms of
the Jacobi elliptic integrals so that it is not easy to find the
explicit form of the dispersion relation. Note that the right hand
side of the dispersion relation in \eq{discirs2} does not depend
on the value of $\r_c$.

\section{More general solution}

In the previous sections, we have considered a solitonic string
configuration (spike or magnon) on either $AdS_3$ or $S^2$. In
this section, we investigate more general solitonic string
configuration expanding on both $AdS_3$ and $S^2$ space. Using
\eq{chrel}, the anomalous dimension for this string configuration
can be rewritten as \bea && E - S - J  \nn &&  = \frac{2 T}{b
(b^2-a^2)} \lb \int_{\r_{min}}^{\r_{max}} d \r \frac{b^2 \o - a
(\pi_0 - \pi_2 - \pi_4) - b^2 \n}{\r'}  + \int_{\th_{min}}^{\pi/2}
d \th \frac{b^2 \n \cos^2 \th}{\th'} \rb \nn && =  \frac{2 (b^2 \o
- a (\pi_0 - \pi_2 + \pi_4) - b^2 \n) T}{b (b^2-a^2)}
\int_{\r_{min}}^{\r_{max}}  \frac{d \r }{\r'} + 2 T \cos \th_{min}
. \eea The  anomalous dimension is reduced to \\
i) for $\D >0$ \be
\la{c1} E - S - J  \approx \frac{ \ls b^2 \o - a (\pi_0 - \pi_2 -
\pi_4) -b^2 \n \rs T}{b \sqrt{\D}} \log \ls \frac{4 \sqrt{\D}}{b
\o T} S \rs + 2 T \cos \th_{min}, \ee ii) for $\D = 0$ \be \la{c2}
E - S - J  \approx \frac{\ls b^2 \o - a (\pi_0 - \pi_2 - \pi_4) -
b^2 \n \rs T} {b \sqrt{\G}} \ls \frac{12 \sqrt{\G}}{b \o T}
\rs^{1/3} S^{1/3} + 2 T \cos \th_{min}, \ee and iii) for $\D < 0$
\be \la{c3} E - S - J  \approx \frac{2 \ls b^2 \o - a (\pi_0 -
\pi_2 - \pi_4) -b^2 \n \rs \left| \D \right|} {b^2 \o \G -  2 a
\pi_2  \left| \D \right|} S + 2 T \cos \th_{min}. \ee

Especially, for $a \n = \pi_4$ the string configuration on $S^2$
becomes a magnon, which has an infinite angular momentum $J$ and a
finite angle difference $\D \ps$. In this case, the worldsheet
momentum $p$ equal to the angle difference $\D \ps$ is given by
\be \cos \frac{p}{2} =  \sin \th_{min} . \ee Using this result,
the anomalous dimension is \be E - S - J = \frac{2 \  ( b^2 \o - a
(\pi_0 - \pi_2 ) - (b^2 - a^2) \n ) \ T}{b (b^2-a^2)}
\int_{\r_{min}}^{\r_{max}}  \frac{d \r }{\r'} + 2 T \sin
\frac{p}{2} , \ee where the integral with respect to $\r$ is
proportional to i) $\log S$ for $\D>0$, ii) $S^{1/3}$ for $\D =0$
and iii) $S$ for $\D < 0$. If $b^2 \o = a (\pi_0 - \pi_2 ) + (b^2
- a^2) \n$, the dispersion relation for this configuration becomes
to that of the magnon on $S^2$ \be E - S - J = 2 T \sin
\frac{p}{2} . \ee Though the above dispersion relation is the same
as that for the circular string located at $\r=\r_c$ in the
section 3, the string configuration considered here is a more
general one extended in the $\r$ direction.

When $a \pi_4 = b^2 \n$, the string configuration on $S^2$ becomes
spike, which has a finite angular momentum and an infinite angle
difference. The difference of the conserved charges becomes \be E
- S - J = \frac{2 \ (b^2 \o - a (\pi_0 - \pi_2 ) ) \ T}{b
(b^2-a^2)} \int_{\r_{min}}^{\r_{max}}  \frac{d \r }{\r'} + 2 T
\cos \th_{min} . \ee The dispersion relation for the spike
configuration is given by \bea E - S - T \D \ps &=& \frac{2 \ (b^2
\o - a (\pi_0 - \pi_2 ) + b(b^2-a^2) \n /a ) \ T}{b (b^2-a^2)}
\int_{\r_{min}}^{\r_{max}}  \frac{d \r }{\r'} \nn
&&  + 2 T \td{\th} %\frac{2 b T}{a} \frac{\td{\th}}{\cos \td{\th}},
\eea where $\td{\th} = \pi/2 - \th_{min}$ and the integral with
respect to $\r$ splits to the three cases like \eq{c1}, \eq{c2}
and \eq{c3} depending on the value of $\D$. For $a (\pi_0 - \pi_2
) = b^2 \o + b(b^2-a^2) \n /a $, this dispersion relation looks
like one for a spike on $S^2$.

\section{Conclusions}

In this paper, we have investigated a class of string
configurations on $AdS_3 \times S^2$ with a particular parameter
relation $k=\o$.
%In general, though there exist various string
%configurations for $\k \ne \o$ we have concentrated on the simpler
%case with $k=\o$.
%In Ref. \ct{Lee:2008sk}, the string
%configuration on $AdS_3 \times S^3$ with some constraints for $\k
%\ne \o$ has been investigated.

For $k=\o$, there are three kinds of the string configuration on
the $AdS_3$ space. Each of these configurations has a different
anomalous dimension proportional to $\log S$ for $\D >0$ and
$S^{1/3}$ for $\D=0$. In these cases, the string is extended from
the minimum of $\r$ to infinity and depending on the position of
$\r_c$ where the string slope ${\cal R}$ diverges, the string
configurations can be split into several cases by studying the
shape of them. For $\D < 0$ since there exist two extremum points
corresponding to minimum and maximum of $\r$, the range of $\r$ is
given by $\r_{min} \le \r \le \r_{max}$, in which the anomalous
dimension is proportional to $S$.

Furthermore, these string configurations have been generalized to
the $AdS_3 \times S^2$ case having two angular momenta, $S$ in the
$AdS$ and $J$ in the $S^2$. Here, we have obtained the generalized
dispersion relation for the string configuration on $AdS_3 \times
S^2$.

\vskip .2in \noindent {\bf Acknowledgements:} B.-H. Lee and B. Gwak
were supported by the Science Research Center Program of the
Korean Science and Engineering Foundation through the Center for
Quantum SpaceTime (CQUeST) of Sogang University with grant number
R11-2005-021. C. Park was supported by the Korea Research Council
of Fundamental Science and Technology (KRCF).

%%%%%%%%%%%%%%%%%%%%%%%%%%%%%%%%%%%%%%
%%%%%%% Thebibligraphy %%%%%%%%%%%%%%%%%%%%%
%%%%%%%%%%%%%%%%%%%%%%%%%%%%%%%%%%%%%

%%%%%%%%%%%%%%%%% Journal Macros %%%%%%%%%%%%%%%%%%%%%%%%%%%
\newcommand{\np}[3]{Nucl. Phys. {\bf B#1}, #2 (#3)}
\newcommand{\pprd}[3]{Phys. Rev. {\bf D#1}, #2 (#3)}
\newcommand{\jjhep}[3]{J. High Energy Phys. {\bf #1}, #2 (#3)}
%\newcommand{\hep}[1]{{\tt hep-th/{#1}}}
%\newcommand{\app}[3]{Ann. Phys. {\bf #1}, #2, (#3)}
%\newcommand{\prp}[3]{Phys. Rept. {\bf #1}, #2, (#3)}
%\newcommand{\jmp}[3]{J. Math. Phys. {\bf #1}, #2, (#3)}
%%%%%%%%%%%%%%%%%%%%%%%%%%%%%%%%%%%%%%%%%%%%%%%%%%%%%%%%%%%%%%

\end{document}